\documentclass[pra, twocolumn,amsfonts,amsmath,amssymb,eufrak]{revtex4-2}

\usepackage{epsfig}
\usepackage{color}
\usepackage{bm}
\usepackage{hyperref}

\usepackage{braket}
\usepackage{graphicx}
\usepackage{amsthm}

\newcommand{\eq}[1]{\begin{equation} #1 \end{equation}}
\newcommand{\eqa}[2]{\begin{equation} #1 \label{#2} \end{equation}}
\newcommand{\balign}[1]{\begin{align} #1 \end{align}}

\newcommand{\bs}{\boldsymbol}

\newcommand{\figin}[4]
{\begin{figure}[tb]
\centering
\includegraphics[width= #1]{#2.pdf}
\caption{#3}
\label{f:#4}
\end{figure}}

\newcommand{\todayd}{\the\year/\the\month/\the\day}

\newcommand{\bib}{\bibitem}

\newcommand{\up}{\uparrow}
\newcommand{\down}{\downarrow}

\newcommand{\lr}{\leftrightarrow}

\newcommand{\lb}{\label}
\newcommand{\nt}{\notag}
\newcommand{\Tr}{\mathrm{Tr}}

\newcommand{\bel}{\begin{easylist}}
\newcommand{\eel}{\end{easylist}}

\newcommand{\eref}[1]{Eq.~\eqref{#1}}
\newcommand{\tref}[1]{Theorem.~\ref{t:#1}}

\newcommand{\lref}[1]{Lemma.~\ref{t:#1}}

\newcommand{\fref}[1]{Fig.~\ref{f:#1}}

\def \({\left(}
\def \){\right)}
\def \[{\left[}
\def \]{\right]}

\newcommand{\abs}[1]{\left|#1\right|}

\newcommand{\sumtwo}[2]%
{\mathop{\sum_{#1}}_{#2}}
\newcommand{\sumthree}[3]%
{\mathop{\mathop{\sum_{#1}}_{#2}}_{#3}}
\newcommand{\sumfour}[4]%
{\mathop{\mathop{\mathop{\sum_{#1}}_{#2}}_{#3}}_{#4}} 
\newcommand{\prodtwo}[2]%
{\mathop{\prod_{#1}}_{#2}}
\newcommand{\mintwo}[2]%
{\mathop{\min_{#1}}_{#2}}
\newcommand{\maxtwo}[2]%
{\mathop{\max_{#1}}_{#2}}
\newcommand{\maxthree}[3]%
{\mathop{\mathop{\max_{#1}}_{#2}}_{#3}}
\newcommand{\limtwo}[2]%
{\mathop{\lim_{#1}}_{#2}}
\newcommand{\suptwo}[2]%
{\mathop{\sup_{#1}}_{#2}}
\newcommand{\supthree}[3]%
{\mathop{\mathop{\sup_{#1}}_{#2}}_{#3}}
\newcommand{\supfour}[4]%
{\mathop{\mathop{\mathop{\sup_{#1}}_{#2}}_{#3}}_{#4}} 
\newcommand{\inftwo}[2]%
{\mathop{\inf_{#1}}_{#2}}
\newcommand{\infthree}[3]%
{\mathop{\mathop{\inf_{#1}}_{#2}}_{#3}}
\newcommand{\inffour}[4]%
{\mathop{\mathop{\mathop{\inf_{#1}}_{#2}}_{#3}}_{#4}} 

\newcommand\calG{{\cal G}}

\newcommand\calN{{\cal N}}

\newcommand\calS{{\cal S}}

\newcommand{\bsxi}{{\boldsymbol{\xi}}}

\newcommand{\bbR}{\mathbb{R}}

\newcommand{\ep}{\varepsilon}

\newcommand{\para}[1]{{\em #1}\/.---}

\newtheorem{thm}{Theorem}
\newtheorem{lm}[thm]{Lemma}
\newtheorem{pro}[thm]{Proposition}

\newcommand{\bthm}[1]{\begin{thm} #1 \end{thm}}
\newcommand{\blm}[1]{\begin{lm} #1 \end{lm}}

\theoremstyle{definition}
\newtheorem{dfn}[thm]{Definition}

\def\rnum#1{\resizebox{0.5em}{\height}{\expandafter{\romannumeral #1}}}
\def\Rnum#1{\resizebox{0.5em}{\height}{\uppercase\expandafter{\romannumeral #1}}}

\newcommand{\HSK}{H_{\rm SK}}
\newcommand{\Hcl}{H_{\rm cl}}
\newcommand{\fSK}{f_{\rm SK}}
\newcommand{\fREM}{f_{\rm REM}}
\newcommand{\fcl}{f_{\rm cl}}

\begin{document}


\title{
Exactly solvable quantum spin-glass model with 1RSB-fullRSB transition
}

\author{Naoto Shiraishi} 
\email{shiraishi@phys.c.u-tokyo.ac.jp}
\affiliation{Department of Basic Science, The University of Tokyo, 3-8-1 Komaba, Meguro-ku, Tokyo 153-8902, Japan}%



\begin{abstract}
We introduce a novel quantum spin-glass model, a Sherrington-Kirkpatrick model with a transverse mean-field type random magnet.
We rigorously derive the exact expression of the free energy of this model at the entire parameter region.
The obtained exact solution implies the existence of a 1RSB-fullRSB transition at low temperatures.
Our technique can be applied to general classical spin models, telling that any solvable classical spin model has its solvable quantum counterpart.
\end{abstract}

\maketitle

\para{Introduction}
Spin systems with quenched randomness have been studied as models of various systems from disordered materials to combinatorial optimization problems~\cite{Nisbook, MMbook}.
In particular, the Sherrington-Kirkpatrick (SK) model~\cite{SK75} serves as a prototype of disordered spin systems.
The SK model is a full-connected spin model whose interaction coefficients are randomly quenched.
In spite of its simplicity, this model possesses various key properties of disordered systems including the replica symmetry breaking and the ultrametricity, which explains why the SK model is considered to be a representative of spin-glass systems.
The famous Parisi formula, which provides the free energy of the SK model, was first proposed on the basis of the replica trick~\cite{Par79, Par80a} and finally justified with a rigorous mathematical proof~\cite{Gue03, ASS03, Tal06b, Talbook, Panbook}.

Recently, quantum disordered spin models have attracted interest from various research fields.
One interest comes from the investigation of quantum many-body localization phenomena~\cite{BAA06, NH15}, where a system does not thermalize and is frozen near its initial state due to quenched randomness.
Some quantum spin-glass models are considered good stages to examine localization~\cite{LPS14, Bal16, Bal17}.
Another interest comes from the field of quantum computation, more specifically, the quantum annealing~\cite{KN98} or the quantum adiabatic computation~\cite{Far00, AL18}, where some hard combinatorial optimization problems are computed by using quantum effects.
One bottleneck of quantum annealing is considered to be the existence of the spin-glass phase~\cite{Jor08, Jor10, Bap13, Kny16, You17, MRC18, MC19, YSH24}.
Of course, the theoretical understanding of quantum effects in disordered spin systems itself, i.e.,  clarifying how robust properties in classical disordered systems are in a quantum regime, is worth investigating deeply~\cite{FS86, YI87, RCC89, GL90, Gol90, NR98, ONS07, Tak07, Bir21}.
These backgrounds motivate the study of quantum disordered spin models by a variety of approaches including numerical simulations~\cite{Bal16}, replica analyses~\cite{Gol90, ONS07, LPS14, Bal16}, and rigorous mathematics~\cite{Cra07, AB19, MW20, Los21, MW23, MW24}.

To take the quantum effect into account, the SK model with a transverse quantum magnetic field has been frequently studied.
Its behavior was studied first by numerical simulations~\cite{RCC89, You17, MRC18, MC19} and the replica method~\cite{FS86, YI87, GL90, Tak07}, and then some results are rigorously founded~\cite{Cra07, AB19, Los21, MW24}.
In particular, the existence of the replica symmetry breaking in the quantum regime is demonstrated in a narrow but finite parameter region~\cite{Los21}.
However, compared to classical spin-glass models, rigorous analyses on quantum spin-glass models are still limited, and most results are based on numerical simulations, the replica method, or other approximations.
This suggests the necessity of a more tractable quantum spin-glass models.

In this Letter, we propose a novel type of quantum spin-glass model, the SK model with a transverse mean-field type random magnet.
This additional term is essentially the same as the random energy model (REM)~\cite{Der81}, which is an extremely simplified model of disordered spin systems.
The energies of REM are quenched randomly and independently, for which the REM is sometimes regarded as a mean-field spin-glass model~\cite{OP84, Bal16}.
The REM is solvable and shows one-step replica symmetry breaking (1RSB)~\cite{Der81, MMbook, Bovbook}.
Our model has both a longitudinal random spin interaction (the SK model) and a transverse random spin interaction (the REM), both of which solely have different types of RSB phases, the full replica symmetry breaking (fullRSB) phase and the 1RSB phase.
Remarkably, in spite of the apparent complexity of this model, the free energy of this model is exactly and rigorously solvable and its phase diagram can be calculated.
In particular, this model accompanies a 1RSB-fullRSB transition at low temperatures.
Compared to other known rigorous results on the quantum SK models, the proof is much simpler and easier to handle.

The proposed proof technique applies not only to the quantum SK model but to all quantum models constructed from classical spin systems both with and without quenched randomness.
We demonstrate that the quantum model constructed by adding a transverse mean-field type random magnet to a classical spin model is exactly solvable in the sense that its free energy can be expressed by using only the free energy of the corresponding classical spin model.
This result implies a striking fact that all solvable classical spin models have their solvable quantum counterparts.

\para{Solvable quantum spin-glass model}
We consider an $N$ spin system of $S=1/2$ with quenched randomness.
The system Hamiltonian is a composition of the classical Sherrington-Kirkpatrick (SK) model in the $z$-axis and the transverse random energy field in the $x$-axis. 
The Hamiltonian of the classical SK model is given by
\eq{
\HSK^z:=\frac{1}{\sqrt{N}}\sum_{i<j}J_{ij}S_i^z S_j^z,
}
where the coefficient $J_{ij}$ is quenched random variables drawn from the standard Gaussian distribution $\calN(0,1)$.

Most of previous investigations of quantum SK models add the $x$ magnetic field as the quantum term~\cite{RCC89, You17, MRC18, MC19, FS86, YI87, GL90, Tak07, Cra07, AB19, Los21, MW24}.
Instead, in this Letter, we add a mean-field type random quantum term, the random energy field in the $x$-axis.
The transverse random energy field is given by~\cite{Der81}
\eq{
R^x=\sum_{\bsxi\in \{ \pm\}^{\otimes N}}  E_{\bsxi}\ket{\bsxi^x}\bra{\bsxi^x},
}
where $\bsxi$ runs all possible $2^N$ sequences of $+$ and $-$ with length $N$, and $\ket{\pm}:=\frac{1}{\sqrt{2}}(\ket{\up}\pm\ket{\down})$.
The energy $E_{\bsxi}$ is quenched random variables drawn from the rescaled Gaussian distribution $\calN(0,N)$ whose variance is $N$.
Compared to random Hamiltonian perturbation studied in the context of thermalization phenomena~\cite{DR20, Roo24, Tas24}, this perturbation retains a physical structure that the eigenstates are aligned to the $x$-axis.

The total quantum Hamiltonian is defined as
\eqa{
H(\Gamma)=\HSK^z+\Gamma R^x
}{Htot}
with nonnegative parameter $\Gamma\geq 0$.
A small $\Gamma$ means a small quantum effect, implying that the Hamiltonian approaches the original SK Hamiltonian.

When the Hamiltonian has quenched randomness, the free energy density $f(\beta)=\frac{1}{\beta N}\ln Z=\frac{1}{\beta N}\ln \Tr[e^{-\beta H}]$ is also a random variable.
However, it is known that both the free energy densities of the SK model $\HSK$ and the random energy model $R$ satisfy self-averaging, i.e., a certain value is realized almost surely~\cite{PS91}.
We express these two free energy densities for inverse temperature $\beta$ as $\fSK(\beta)$ and $\fREM(\beta)$, respectively.
These two free energy densities are solved exactly~\cite{Der81, Tal06b, Talbook, Panbook}.

Our subject is the free energy density for $H(\Gamma)$.
Although this is a random variable, as we show later, the free energy density almost surely takes a certain value.
We denote this value by $f(\beta, \Gamma)$, which is realized in almost all realizations of quenched randomness.
We claim that $f(\beta, \Gamma)$ is exactly solvable for all $\beta$ and $\Gamma$.
\bthm{\lb{t:main}
The free energy density of the system with Hamiltonian \eqref{Htot} almost surely exists in the thermodynamic limit and satisfies
\eqa{
f(\beta, \Gamma)=\max\[ \fSK(\beta), \Gamma \fREM(\Gamma\beta)\].
}{main}
}
Since both $\fSK$ and $\fREM$ are already computed in previous studies, the above solution means the exact solvability of the quantum SK model $H(\Gamma)$ in the entire parameter space.
The known free energy expression of a quantum SK model~\cite{MW24} accompanies a complicated path integral or uneasy optimization, which prohibits a transparent understanding of the phase diagram.
Compared to this, the free energy of our quantum SK model has a simple and tractable expression as long as the free energy of the classical SK model has already been calculated.

At sufficiently low temperatures, the SK model is in the fullRSB phase, and the REM is in the 1RSB phase.
Hence, by changing $\Gamma$ at such a low temperature, we observe a phase transition from the 1RSB phase to the fullRSB phase.
This 1RSB-fullRSB transition in the quantum regime is rigorously shown in our model.

\figin{8.5cm}{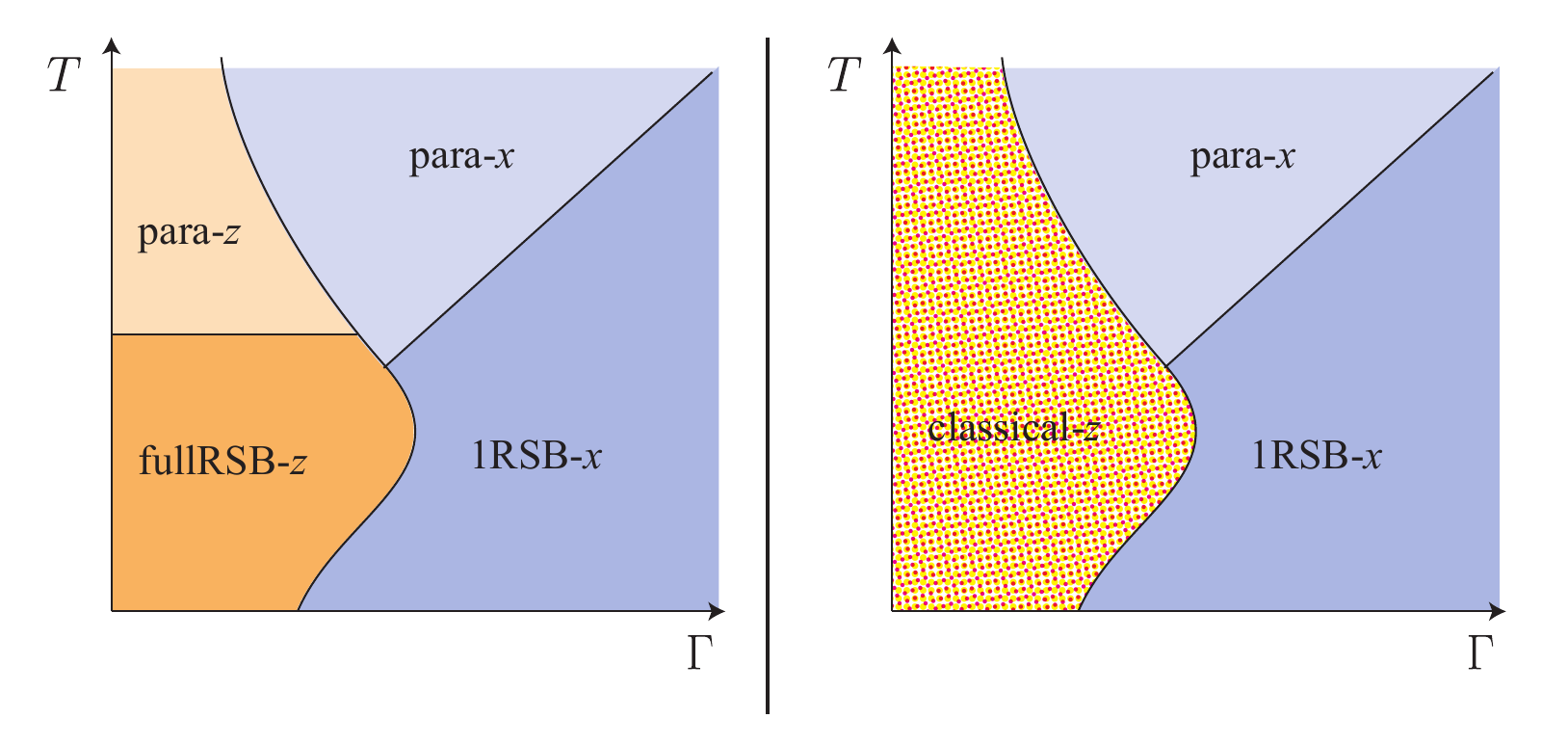}{
(Left): An example of the phase space where the classical spin system is an SK-type model. 
We have a 1RSB-fullRSB transition at low temperatures.
(Right): The structure of the phase space for the Hamiltonian \eqref{H-gen}.
The small $\Gamma$ region fully depends on the details of the classical Hamiltonian, while the large $\Gamma$ region is irrelevant to the classical Hamiltonian.
}{phase}

\para{Extension to general spin systems}
As will be seen, the proof of \tref{main} does not rely on the specialty of the SK model.
In fact, below we claim that for any classical spin system, a counterpart quantum model constructed by adding a transverse mean-field type random magnet (random energy field) is exactly solvable, and its free energy takes a form similar to \eref{main}.

Consider a general classical $N$ spin system with at most $k$-body interaction.
A single spin takes two values, $\up$ or $\down$, and a spin operator $S$ takes $+1$ (resp. $-1$) if the spin is $\up$ (resp. $\down$).
Since any Boolean function $G:\{\pm 1\}^{\otimes m}\to \bbR$ can be expanded by products of spin operators $S_{i_1}S_{i_2}\cdots S_{i_{m'}}$ with $m'\leq m$~\cite{ODbook}, without loss of generality we restrict the system Hamiltonian in the following form
\eqa{
\Hcl=\sum_{m=1}^k \sum_{1\leq i_1<i_2<\cdots < i_m\leq N}a_{i_1,\ldots , i_m}S_{i_1}\cdots S_{i_m},
}{genH}
where $i_1,\ldots , i_m$ run all possible combinations of $m$ elements in $\{1,\ldots , N\}$ in the increasing order.
We suppose that the coefficients $a_{i_1,\ldots , i_m}$ are not extremely large.
Precisely, we require that there exists a real number $b$ such that 
\eqa{
\lim_{N\to \infty}\frac{\max_{1\leq m\leq k}\max_{i_1,\ldots,i_m}a_{i_1,\ldots , i_m}}{(\ln N)^b}=0,
}{genH-cond}
which confirms that the interaction coefficients are at most a polynomial of $\ln N$.
We now state our second main result on general classical spin systems.

\bthm{\lb{t:gen}
Consider a spin system with Hamiltonian \eqref{genH} with condition \eqref{genH-cond}, whose free energy density $\fcl$ uniquely exists in the thermodynamic limit.
Then, the corresponding quantum system with Hamiltonian
\eqa{
H=\Hcl^z+\Gamma R^x
}{H-gen}
has the free energy density given by
\eqa{
f(\beta, \Gamma)=\max[ \fcl(\beta), \Gamma\fREM(\Gamma\beta)].
}{gen-main}
}

This theorem claims a striking fact that any solvable classical spin system has a solvable quantum counterpart with the phase diagram as \fref{phase}.
The classical Hamiltonian $\Hcl$ can be both deterministic and probabilistic.
For example, taking the two-dimensional ferromagnetic Ising model~\cite{Ons44} as $\Hcl$, we get a two-dimensional exactly-solvable quantum spin system.

In various contexts of research, whether a property seen in classical systems is retained under quantum perturbation is a subject of debate.
Surprisingly, our theorem comprehensively solves this type of general problem in the affirmative as long as the property in interest is determined by the free energy.
We consider that the REM-type quantum effect is the simplest way to take the quantum effect into account, where only two extremal regimes are picked up and intermediate regimes are suppressed.

\para{Proof of \tref{main}: lower bound}
Our proof of \tref{main} is inspired by the technique presented in Ref.~\cite{MW20}.
Let $f_N(\Gamma, \beta)$ be the free energy density of $H(\Gamma)$ at inverse temperature $\beta$ with $N$ spins.
First, we shall show that the following lower bounds hold almost surely:
\balign{
\liminf_{N\to \infty}f_N(\beta, \Gamma)\geq& \Gamma \fREM(\Gamma\beta), \lb{lower1} \\
\liminf_{N\to \infty}f_N(\beta, \Gamma)\geq& \fSK(\beta). \lb{lower2}
}
To prove Eqs.~\eqref{lower1} and \eqref{lower2}, we use the Gibbs variational principle for any density matrix $\rho$:
\eqa{
\ln \Tr[e^{-\beta H(\Gamma)}]\geq -\beta \Tr[H(\Gamma)\rho]+\Tr[\rho \ln \rho].
}{Gvp}

Observe that $\HSK^z\ket{\bsxi}$ with $\bsxi\in \{\pm\}^{\otimes N}$ is a superposition of states where two spins are flipped from $\ket{\bsxi}$ as $+\lr -$, and hence $\braket{\bsxi|\HSK^z|\bsxi}=0$ for any $\bsxi\in \{\pm\}^{\otimes N}$.
Then, \eref{lower1} follows from \eref{Gvp} by setting $\rho=\rho_1=e^{-\Gamma \beta R^x}/\Tr[e^{-\Gamma \beta R^x}]$, where we used $\Tr[\HSK^z\rho_1]=0$.

To derive \eref{lower2}, we set $\rho=\rho_2=e^{- \beta \HSK^z}/\Tr[e^{-\beta \HSK^z}]$ in \eref{Gvp}, which reads 
\eqa{
f_N(\beta, \Gamma)=\frac1N \ln \Tr[e^{-\beta H(\Gamma)}]\geq \fSK(\beta)-\frac{\beta}{N} \Tr[R^x \rho_2].
}{lower2-mid}
Noticing that any ${\bs \alpha}\in \{ \up, \down\}^{\otimes N}$ and $\bsxi\in \{\pm\}^{\otimes N}$ satisfy $\abs{\braket{\bs \alpha|\bsxi}}^2=\frac{1}{2^N}$, we compute $\Tr[R^x \rho]$ as
\balign{
\Tr[R^x \rho]=&\sum_{\substack{{\bs \alpha}\in \{ \up, \down\}^{\otimes N}, \\ \bsxi\in \{\pm\}^{\otimes N}}}\abs{\braket{\bs \alpha|\bsxi}}^2E_{\bsxi}\braket{\bs \alpha|\rho|\bs \alpha}=\frac{1}{2^N}\sum_{\bsxi\in \{\pm\}^{\otimes N}}E_{\bsxi}. \lb{lower2-mid2}
}
Since $2^N$ elements of $E_{\bsxi}$ are drawn independently from a rescaled Gaussian distribution $\calN(0,N)$, the right-hand side of \eref{lower2-mid2} is a stochastic variable drawn from a rescaled Gaussian distribution $\calN(0,2^{-N}N)$, which is almost surely sublinear in $N$.
We thus conclude that the second term of the right-hand side of \eref{lower2-mid} vanishes almost surely, and the desired relation \eqref{lower2} is obtained.

\para{Proof of \tref{main}: upper bound}
We next derive the following upper bound:
\eqa{
\limsup_{N\to \infty}f_N(\beta, \Gamma)\leq \max(\fSK(\beta), \Gamma \fREM(\Gamma\beta)).
}{upper}
Combining Eqs.~\eqref{lower1}, \eqref{lower2}, and \eqref{upper}, we obtain the desired result \eqref{main}.

To derive \eref{upper}, we introduce a low-energy subspace of $\{\pm\}^{\otimes N}$ with respect to $R^x$:
\eqa{
\calS_\ep:=\{ \bsxi \in \{\pm\}^{\otimes N}|E_\bsxi\leq -\ep N\} .
}{def-Sep}
We decompose $\calS_\ep$ into several groups according to the criteria that two states $\bsxi$ and $\bsxi'$ {\it neighbor} if the Hamming distance between these two, denoted by $d(\bsxi, \bsxi')$, is less than or equal to 4 (see \fref{image-C}).
More precisely, $\calS_\ep$ is decomposed into groups $C_\ep^{m}$ such that (i) for any two states in different groups $\bsxi\in C_\ep^m$ and $\bsxi'\in C_\ep^{m'}$ with $m\neq m'$, $d(\bsxi, \bsxi')>4$ holds, (ii) for any two states in the same group $\bsxi, \bsxi'\in C_\ep^m$, there exists a sequence $\bsxi=\bsxi_1, \bsxi_2, \ldots, \bsxi_n= \bsxi'$ with $\bsxi_i\in C_\ep^m$ and $d(\bsxi_i, \bsxi_{i+1})\leq 4$ for all $i$.
As guessed, the maximum size of $C_\ep^m$ is not so large.
In fact, we can prove that 
\eqa{
\max \abs{C_\ep^m} \leq M_\ep =O(\ep^{-2})
}{maxC-bound}
holds almost surely, where $M_\ep$ is a quantity independent of $N$ (see \lref{gen-D-bound} in the End Matter).

\figin{8.5cm}{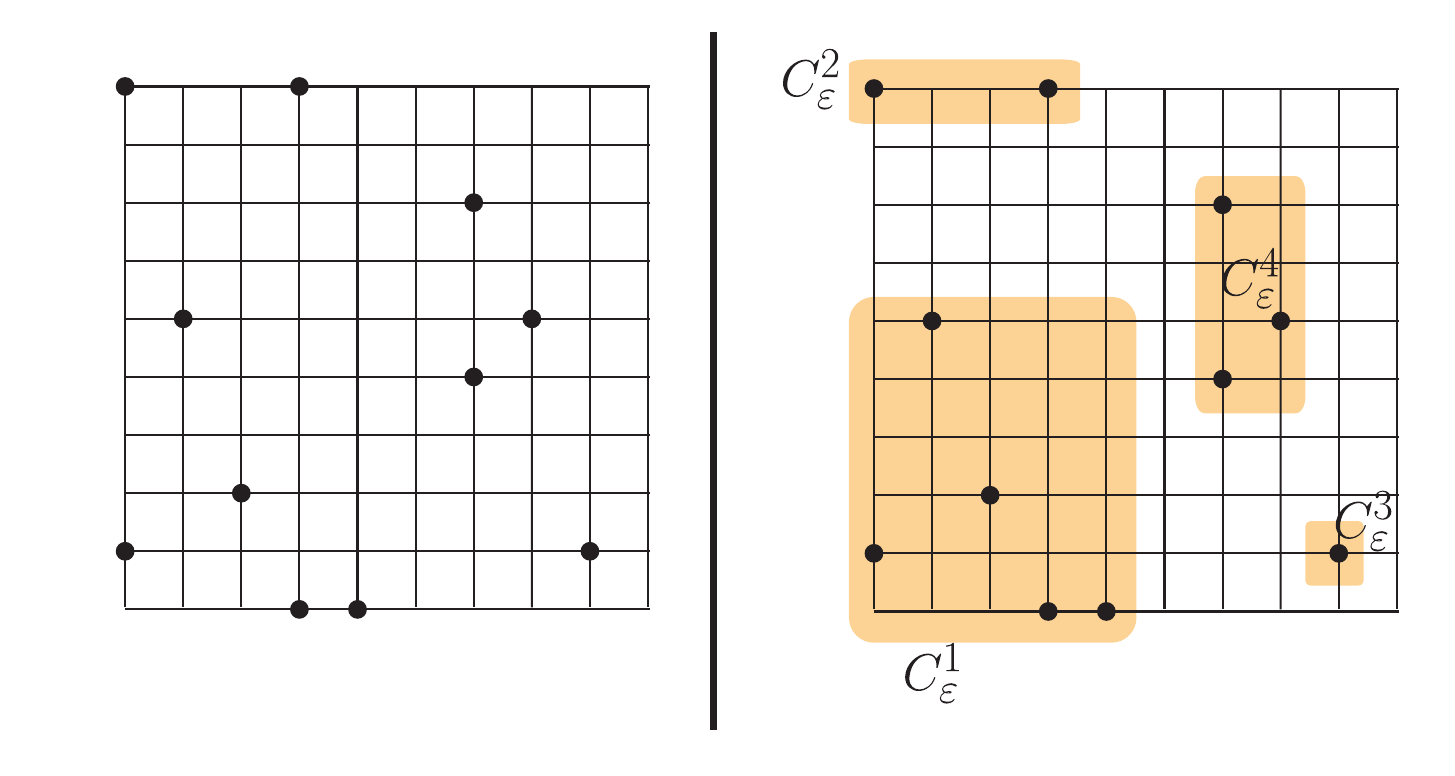}{
An example of how $C_\ep^m$ is determined.
Black points represent low energy states in $\calS_\ep$.
Remark that the correct state space is corners of a high-dimensional hypercube, not a large two-dimensional lattice.
}{image-C}

Correspondingly, we decompose the Hamiltonian $H(\Gamma)$ into two parts, $H^1$ and $H^2$.
To this end, we employ the basis $\{ \ket{\bsxi}\}$ with $\bsxi \in \{\pm\}^{\otimes N}$ and represent $H$, $\HSK^z$, and $R^x$ as $N\times N$ matrices in this basis.
We decompose $\HSK^z$ into $(\HSK^z)_{\bsxi, \bsxi'}$ with $\bsxi\in \calS_\ep$ or $\bsxi'\in \calS_\ep$ and that with $\bsxi\notin \calS_\ep$ and $\bsxi'\notin \calS_\ep$, which we denote by $\HSK^1$ and $\HSK^2$, respectively.
Using them, we decompose $H(\Gamma)$ into $H^1:=\HSK^1$ and $H^2:=\HSK^2+\Gamma R^x$.

$\HSK^1$ is further decomposed as $\HSK^1=\sum_m \HSK^{C_\ep^m}$, where $\HSK^{C_\ep^m}$ consists of matrix elements $(\HSK^z)_{\bsxi, \bsxi'}$ for $\bsxi\in C_\ep^m$ or $\bsxi'\in C_\ep^m$.
Since different $C$'s (i.e., $C_\ep^m$ and $C_\ep^{m'}$ with $m\neq m'$) are distant from each other by Hamming distance larger than 4, this is a well-defined decomposition of $\HSK^1$, and off-diagonal blocks of $\HSK^1$ with different $C$'s vanish (i.e., $\HSK^{C_\ep^m}\HSK^{C_\ep^{m'}}=0$ for $m\neq m'$).
Using this decomposition, the operator norm of $H^1=\HSK^1$ is bounded from above as
\balign{
\| H^1\|=\max_m \| \HSK^{C_\ep^m}\| &\leq \max_m\sqrt{\sum_{\bsxi, \bsxi'}\abs{\braket{\bsxi|\HSK^{C_\ep^m}|\bsxi'} }^2} \nt \\
&\leq \sqrt{2N\max_{i,j}\frac{J_{ij}^2}{N} \max_m \abs{C_m^\ep}^2} .
}
The first equality comes from the absence of off-diagonal blocks with different $C$'s.
In the last inequality, we used the fact that $\braket{\bsxi|\HSK^{C_\ep^m}|\bsxi'} $ takes $J_{ij}$ if $\bsxi$ and $\bsxi'$ differs only sites $i$ and $j$ and takes zero otherwise.
Since $\max_{i,j}\abs{J_{ij}}\leq \sqrt{6\ln N}$ holds almost surely (see End Matter), we have the following upper bound of the operator norm of $H^1$:
\eq{
\|H^1\|=\|\HSK^1\|\leq 2\sqrt{3\ln N}\max_m \abs{C_m^\ep}.
}


Now we compute the partition function $Z(\beta, \Gamma)$.
Using the Golden-Thompson inequality $\Tr[e^{A+B}]\leq \Tr[e^Ae^B]$ for Hermitian matrices $A$ and $B$, we have
\eq{
Z(\beta, \Gamma)\leq \Tr[e^{-\beta H^1}e^{-\beta H^2}]\leq e^{\beta \|H^1\|}\Tr[e^{-\beta H^2}].
}
The second term of the right-hand side can be split into contributions from $\calS_\ep$ and $\calS_\ep^{\rm c}$ as
\eqa{
\Tr[e^{-\beta H^2}]=\Tr_{\calS_\ep}[e^{-\beta \Gamma R^x_{\calS_\ep}}]+\Tr_{\calS_\ep^{\rm c}}[e^{-\beta (\HSK^2+ \Gamma R^x_{\calS_\ep^{\rm c}})}],
}{upper-mid1}
where $\Tr_{\calS_\ep}$ and $\Tr_{\calS_\ep^{\rm c}}$ are traces whose ranges are $\calS_\ep$ and $\calS_\ep^{\rm c}$, and $R^x_{\calS_\ep}$ and $R^x_{\calS_\ep^{\rm c}}$ are restrictions of $R^x$ to $\calS_\ep$ and $\calS_\ep^{\rm c}$,  respectively.

Since $R^x$ is diagonalized in the basis $\{\ket{\pm}\}^{\otimes N}$, the first term of the right-hand side of \eref{upper-mid1} is bounded as
\eq{
\Tr_{\calS_\ep}[e^{-\beta \Gamma R^x_{\calS_\ep}}]\leq \Tr[e^{-\beta \Gamma R^x}]=e^{N\Gamma \beta \fREM(\Gamma\beta)}.
}
Using the Golden-Thompson inequality twice, the second term of the right-hand side of \eref{upper-mid1} is bounded as
\balign{
&\Tr_{\calS_\ep^{\rm c}}[e^{-\beta (\HSK^2+ \Gamma R^x_{\calS_\ep^{\rm c}})}]=\Tr_{\calS_\ep^{\rm c}}[e^{-\beta (\HSK- \HSK^1+ \Gamma R^x_{\calS_\ep^{\rm c}})}] \nt \\
&\leq \Tr_{\calS_\ep^{\rm c}}[e^{-\beta \HSK}e^{\beta \HSK^1}e^{-\beta \Gamma R^x_{\calS_\ep^{\rm c}})}] \nt \\
&\leq e^{\beta \|\HSK^1\|}e^{\beta \Gamma \ep N}\Tr_{\calS_\ep^{\rm c}}[e^{-\beta \HSK}].\lb{upper-mid2}
}
In the last inequality, we used the fact that $R^x$ is larger than $-\ep N$ in $\calS_\ep^{\rm c}$.
Since $e^{-\beta \HSK}$ is a nonnegative matrix, the last term of the right-hand side of \eref{upper-mid2} is further evaluated as
\eq{
\Tr_{\calS_\ep^{\rm c}}[e^{-\beta \HSK}]\leq \Tr[e^{-\beta \HSK}]=e^{\beta N \fSK(\beta)}.
}

Combining these relations, we arrive at
\eq{
f(\beta, \Gamma)\leq \max \[\Gamma \fREM(\Gamma\beta), \fSK(\beta)+\ep \Gamma +\frac{\|H^1\|}{N} \] +\frac{\|H^1\|}{N}.
}
Recalling that $\|H^1\|$ is sublinear in $N$ and $\ep$ can be arbitrarily small, we obtain the desired upper bound \eqref{upper}.

\para{Proof of \tref{gen}}
The proof of \tref{gen} is essentially the same as that for \tref{main} with replacing $\HSK$ by $\Hcl$.
In fact, the same lower bounds as Eqs.~\eqref{lower1} and \eqref{lower2} hold with the same proof by replacing $\HSK$ by $\Hcl$.
By contrast, the same upper bound as \eref{upper} holds with replacing $\HSK$ by $\Hcl$ but its proof needs some modification.
We modify the definition of $C_\ep^m$ such that $\bsxi\in C_\ep^m$ and $\bsxi'\in C_\ep^{m'}$ with $m\neq m'$ implies  $d(\bsxi_i, \bsxi_{i+1})>2k$, and $\bsxi, \bsxi'\in C_\ep^m$ implies a sequence $\bsxi=\bsxi_1, \bsxi_2, \ldots, \bsxi_n= \bsxi'$ with $\bsxi_i\in C_\ep^m$ and $d(\bsxi_i, \bsxi_{i+1})\leq 2k$ for all $i$.
Thanks to \lref{gen-D-bound} shown in the End Matter, under the above modification the bound \eqref{maxC-bound} is still valid with $M_\ep=O(\ep^{-2})$.
The remainder is the same as that for \tref{main}.

\para{Discussion}
We have shown that the quantum spin-glass model whose Hamiltonian is given by \eref{Htot} is exactly solvable in the sense that its free energy is rigorously derived.
The solution tells that there are two regions in the phase space ($T$-$\Gamma$ plane); the SK region in the $z$ direction and the REM region in the $x$ direction, both of which have different types of RSB phases, the 1RSB phase and the fullRSB phase.
At a low temperature, this quantum model shows the 1RSB-fullRSB transition.

An astonishing point of this technique lies in the fact that we can construct solvable quantum counterparts of {\it any} classical spin systems.
In various research contexts, whether a property observed in classical spin models is robust under quantum effects has been asked and investigated.
Surprisingly, this general problem is automatically solved affirmatively by \tref{gen} as long as our interest is a property derived from its free energy.

We remark that although the free energy \eqref{gen-main} takes a very simple form, this system indeed has quantum effects.
The energy eigenstates of \eref{Htot} with small but finite $\Gamma$ are not in the computational basis (i.e., those with $\Gamma=0$) due to the transverse random energy field, which changes the dynamics of this model between the case of finite $\Gamma$ and $\Gamma=0$.
In fact, the quantum REM with a transverse uniform $x$ magnetic field shows a dynamical transition where no anomaly exists in the free energy~\cite{LPS14}.
Clarifying the dynamics of our quantum model is left to future research.

{\it Acknowledgments.} ---
The author thanks Mizuki Yamaguchi and Koji Hukushima for insightful discussion.
The author also thanks Tomoyuki Obuchi for careful reading and helpful comments.
This work was supported by JST ERATO Grant Number JPMJER2302, Japan.


\section*{End Matter}
\subsection*{Demonstration that $\max_{i,j}\abs{J_{ij}}\leq \sqrt{6\ln N}$ holds almost surely}
We here demonstrate that $\max_{i,j}\abs{J_{ij}}$ is almost surely less than $\sqrt{6\ln N}$.
Notice that the complementary error function is bounded as
\balign{
\frac{1}{\sqrt{2\pi}}\int_a^\infty e^{-x^2/2}dx=&\frac{1}{\sqrt{2\pi}}\int_a^\infty e^{[-a^2-(x-a)(x+a)]/2}dx \nt \\
\leq&\frac{1}{\sqrt{2\pi}}\int_a^\infty e^{[-a^2-(x-a)^2]/2}dx=\frac{e^{-a^2/2}}{2}. \lb{cerf-bound}
}
Since $\max_{i,j}\abs{J_{ij}}$ is the largest variable in $N(N-1)/2$ standard Gaussian variables, the probability that $\max_{i,j}\abs{J_{ij}}$ is larger than $6\ln N$ is evaluated as
\eq{
{\rm Prob}[\max_{i,j}\abs{J_{ij}}\leq \sqrt{6\ln N} ]\geq \( 1-\frac{1}{N^3}\) ^{N(N-1)/2},
}
whose right-hand side converges to 1 as $N\to \infty$.

\subsection*{Derivation of \eref{maxC-bound}}
We here prove a generalized version of \eref{maxC-bound}, which covers setups of \tref{gen}.
The definition of $\calS_\ep$ is the same as \eref{def-Sep}.
For a given natural number $l$, we decompose $\calS_\ep$ into groups $D_\ep^{m}$ such that (i) for any two states in different groups $\bsxi\in D_\ep^m$ and $\bsxi'\in D_\ep^{m'}$ with $m\neq m'$ satisfies $d(\bsxi, \bsxi')>l$, (2) for any two states in the same group $\bsxi, \bsxi'\in D_\ep^m$, there exists a sequence $\bsxi=\bsxi_1, \bsxi_2, \ldots, \bsxi_n= \bsxi'$ with $\bsxi_i\in D_\ep^m$ and $d(\bsxi_i, \bsxi_{i+1})\leq l$ for all $i$.
The definition $C_\ep^m$ is a special case of $D_\ep^m$ with $l=4$.
Under this setup, we prove the following lemma:
\blm{\lb{t:gen-D-bound}
The maximum size of $D_\ep^m$ is almost surely bounded from above as
\eqa{
\max \abs{D_\ep^m} \leq K_\ep =O(\ep^{-2}),
}{maxD-bound}
where $K_\ep=2\ln 2/\ep^2+1$ is a constant.
}
An important point is that $K_\ep$ is independent of $N$.
This lemma confirms that low-energy states do not concentrate in terms of the Hamming distance.
In the proof of \tref{gen}, we used this Lemma with setting $l=2k$.

In the following, since $l$ is fixed throughout this proof, we drop $l$ dependence for brevity.
For a given $K$, which is finally set to $K_\ep$, we call a subset of $\{ \pm\}^{\otimes N}$ as a {\it good subset} if (a) this subset has $K$ elements, (b) for any two elements of this subset, there exists a sequence $\bsxi=\bsxi_1, \bsxi_2, \ldots, \bsxi_n= \bsxi'$ of elements of the subset such that $d(\bsxi_i, \bsxi_{i+1})\leq l$ holds for all $i$.
We denote the set of good subsets by $\calG_K$, and its size by $\nu(K):=\abs{\calG_K}$.
The event $\max \abs{D_\ep^m}> K$ means that there exists a good subset $T\in \calG_K$ such that any state $\bsxi$ in $T$ has energy less than $-\ep N$.
Thus, using \eref{cerf-bound} and the union bound, we evaluate
\balign{
&{\rm Prob}[\max \abs{D_\ep^m} \geq K] \nt \\
=&{\rm Prob}[ \exists T\in \calG_K \ {\rm s.t.} \ \forall \bsxi \in T, E_\bsxi\leq -\ep N] \nt \\
\leq&\sum_{T\in \calG_K} {\rm Prob}[\forall \bsxi \in T, E_\bsxi\leq -\ep N]\leq \nu(K) \[ \frac12 e^{-\ep^2 N/2}\] ^K. \lb{end-mid1}
}

\figin{8.5cm}{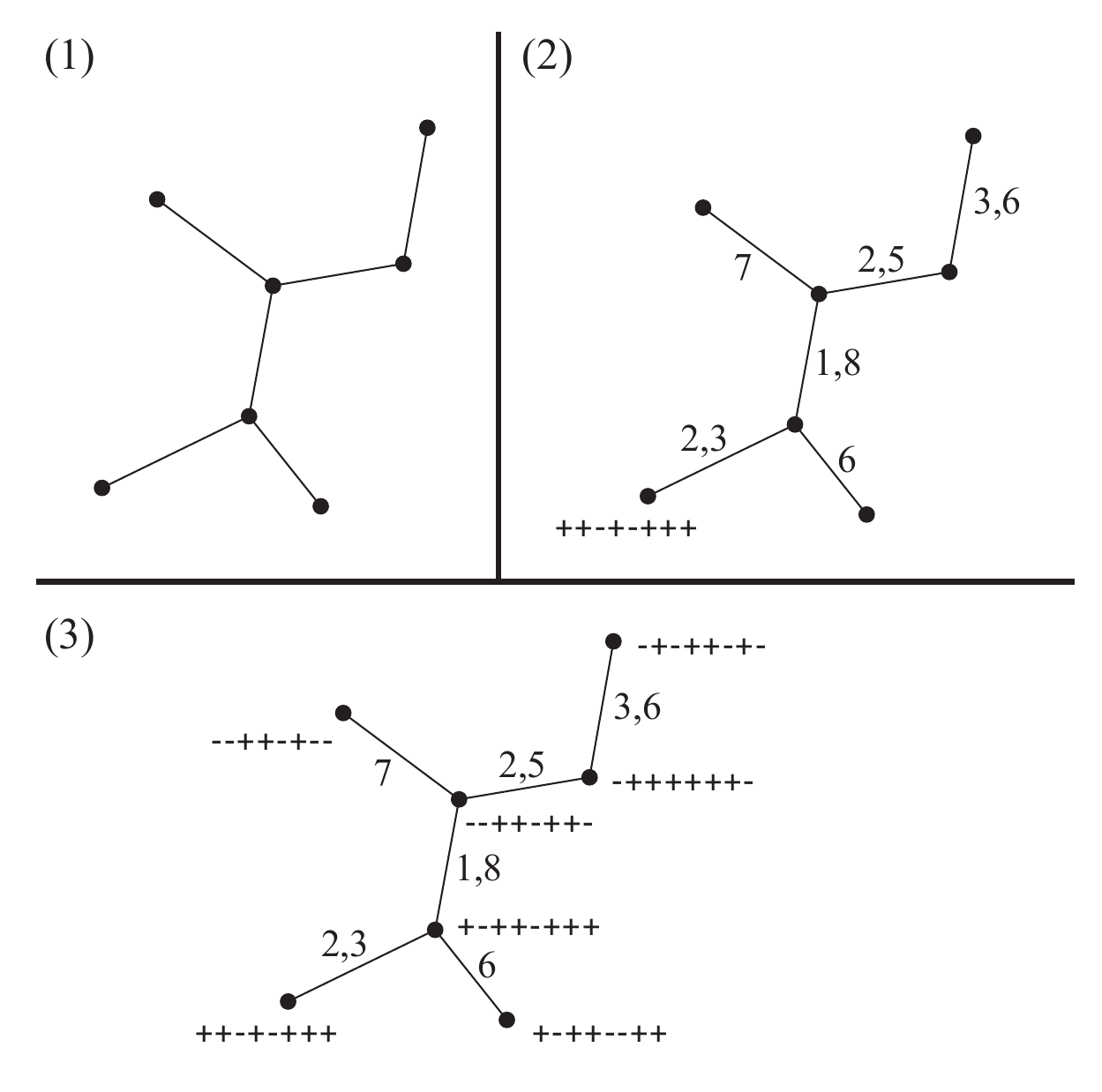}{
Schematics of the evaluation of $\nu(K)$.
We draw an example of $N=8$, $K=7$, and $l=2$.
(1) We prepare a tree graph with $K$ vertices.
(2) We assign at most $l$ numbers from $\{1,\ldots , N\}$.
We also fix a vertex and assign a state.
(3) We assign states to the remaining vertices as follows.
We find an edge whose one end vertex is assigned and the other end vertex is not, and assign a state to the other vertex by flipping sites which are assigned to this edge from the neighboring assigned state.
All good subsets in $\calG_K$ can be obtained by some of these procedures.
}{num-count}

Below we bound $\nu (K)$ from above.
Notice that each good subset $T$ has at least one corresponding tree graph whose vertices are states in $T$ and whose edge connects two states with Hamming distance less than or equal to $l$.
To obtain the tree graph, we first construct a graph by connecting every pair of states whose Hamming distance is less than or equal to $l$ by edge, and if the constructed graph has some loops, we pick up one of the spanning trees.
We note that although a single $T$ may have multiple corresponding tree graphs, a single tree graph satisfying the above conditions has a unique counterpart of a good subset.

To derive an upper bound of $\nu(K)$, we first bound the number of shapes of tree graphs with $K$ vertices denoted by $\mu(K)$, where we do not distinguish each vertex (as (1) in \fref{num-count}).
Notice that any tree graph with $K'+1$ vertices can be obtained by an addition of a vertex to a tree graph with $K'$ vertices, and the added vertex is connected to one of $K'$ vertices.
From this fact, we have $\mu(K'+1)\leq K'\mu(K')$, implying $\mu(K)\leq (K-1)!$.

Next, we assign states to vertices of tree graphs, with satisfying the required condition.
We accomplish this by (i) assigning a state to one fixed vertex, and then (ii) assigning to all edges a set of sites which represents flipped sites between two states on two endpoint vertices ((2) in \fref{num-count}).
Then, all states on all vertices are automatically assigned ((3) in \fref{num-count}).
In (i), we have $2^N$ assignment of states.
In (ii), the assigned set to  a single edge is a subset of $\{ 1,\ldots , N\}$ whose size is at most $l$, whose number is bounded from above by
\eq{
\sum_{i=1}^l \frac{N(N-1)\cdots (N-i+1)}{i!}\leq \sum_{i=1}^l \frac{N^i}{i!}\leq \frac{N^l}{(l-1)!}.
}

Combining our findings, we arrive at the upper bound of $\nu(K)$:
\eqa{
\nu(K)\leq (K-1)! 2^N \( \frac{N^l}{(l-1)!}\) ^{K-1}.
}{end-mid2}
Inserting \eref{end-mid2}, the right-hand side of \eref{end-mid1} reads
\balign{
&(K-1)!2^N \( \frac{N^l}{(l-1)!}\) ^{K-1}\[ e^{-\ep^2 N/2}\] ^K \nt \\
\leq&\exp\[ -N\( \frac{\ep^2 K}{2}-\ln2 \) +K(\ln K +l\ln N-\ln (l-1)!)  \] ,
}
which goes to zero by taking $N\to \infty$ limit with fixed $\ep>0$ and $K>2\ln 2/\ep^2$.
This means that for any $\ep>0$, by setting $K=K_\ep=2\ln 2/\ep^2+1$, then ${\rm Prob}[\max \abs{D_\ep^m} \geq K] $ converges to zero in the $N\to \infty$ limit, which completes the proof of \lref{gen-D-bound}.

\end{document}